\documentclass[11pt]{article}
\usepackage{moriond,epsfig}

\def\be{\begin{equation}}
\def\ee{\end{equation}}
\def\bea{\begin{eqnarray}}
\def\eea{\end{eqnarray}}

\begin{document}
\vspace*{4cm}
\title{Abelian symmetries in multi-Higgs-doublet models}

\author{Igor~P.~Ivanov$^{1,2}$, \underline{Venus~Keus}$^{1}$, Evgeny Vdovin$^{2}$}

\address { $^1$ IFPA, Universit\'{e} de Li\`{e}ge, All\'{e}e du 6 Ao\^{u}t 17, b\^{a}timent B5a, 4000 Li\`{e}ge, Belgium
   \\$^2$ Sobolev Institute of Mathematics, Koptyug avenue 4, 630090, Novosibirsk, Russia
    }

\maketitle\abstracts{
Classifying symmetry groups which can be implemented in the scalar sector of a model with $N$ Higgs doublets is a difficult and an unsolved task for $N>2$.
Here, we make the first step towards this goal by classifying the Abelian symmetry groups. We describe a strategy that identifies all Abelian groups 
which can be realized as symmetry groups of the NHDM scalar potential. 
We give examples of the use of this strategy in 3HDM and 4HDM and prove several statements for arbitrary $N$.}

\section{Introduction}

Many different variants of the Brout-Englert-Higgs (BEH) mechanism of Electroweak symmetry breaking (EWSB) have been proposed so far \cite{CPNSh}, each of which has many free parameters describing the scalar and Yukawa sectors. The standard approach to find a formalism reconstructing the phenomenology of a given model, is to narrow down the freedom by imposing certain symmetries both on the scalar sector and on the Yukawa interactions. However, it is a difficult and often unsolved task to find the full list of symmetries which can be implemented within each variant.

Here we make a step towards solving this problem for models with $N$ Higgs doublets. We focus on Abelian symmetries of the NHDM scalar potential and describe a strategy that algorithmically identifies realizable Abelian symmetry groups for any given $N$. The strategy first addresses groups of unitary transformations in the space of Higgs families and then it is extended to groups that include antiunitary (generalized $CP$) transformations. The strategy also yields explicit examples of the potentials symmetric under a given realizable group. Here we only present the Higgs-family symmetries. 
For the detailed version of the strategy and antiunitary transformations refer to \cite{Ivanov-Keus-Vdovin}.

\section{Symmetries in the scalar sector of NHDM}\label{section-symmetries}

When searching for the symmetry groups which can be implemented in the scalar sector of a non-minimal Higgs model, one must distinguish between {\em realizable} and {\em non-realizable} groups. If it is possible to write a potential which is symmetric under a group $G$ but not symmetric under any larger symmetry group containing it, we call $G$ a realizable group. The true symmetry properties of the potentials are reflected in realizable groups.

We focus on the reparametrization transformations, which are non-degenerate linear transformations mixing different doublets $\phi_a$ but keeping invariant the kinetic term (which includes interaction of the scalar fields with the gauge sector of the model).  

A reparametrization transformation must be unitary (a Higgs-family transformation); $U:  \phi_a \mapsto U_{ab}\phi_b $ or antiunitary (a generalized $CP$-transformation); $U_{CP} = U \cdot J:  \phi_a \mapsto U_{ab}\phi^\dagger_b$, with a unitary matrix $U_{ab}$. The transformation $J \equiv CP$ acts on doublets by complex conjugation and satisfies $J^2 = 1$.

Here we focus on the unitary transformations $U$. A priori, such transformations form the group $U(N)$. However,
the overall phase factor multiplication is already taken into account by the $U(1)_Y$ from the gauge group.
This leaves us with the $SU(N)$ group of reparametrization transformations. Then, this group has a non-trivial center $Z(SU(N))= Z_N$ generated by the diagonal matrix $\exp(2\pi i/N)\cdot 1_N$, where $1_N$ is the identity matrix. These transformations act trivially in the space of potentials. Therefore, the group of {\em physically distinct} unitary reparametrization transformations is the projective special unitary group $G_u = PSU(N) \simeq SU(N)/Z_N$.

\section{Finding Abelian subgroups and identifying the symmetries of the potential}\label{section-strategy}

A maximal Abelian subgroup of $G_u$ is an Abelian group that is not contained
in a larger Abelian subgroup of $G_u$. A priori, there can be several maximal Abelian subgroups in a given group.
Any subgroup of $G_u$ must be either a maximal one, or lie inside a maximal one.
Therefore, we first need to identify all maximal Abelian subgroups of $G_u$ and then study their realizable subgroups.

It was proved in \cite{Ivanov-Keus-Vdovin} that there are two types of maximal abelian subgroups of PSU(N):
maximal tori constructed below and additional finite groups which must be treated separately.
All maximal tori in PSU(N) are conjugate to the group of phase rotations
$T = U(1)_1\times U(1)_2 \times \cdots \times \overline{U(1)}_{N-1}$ where
\bea
U(1)_1 & = & \alpha_1(-1,\, 1,\, 0,\, 0,\, \dots,\, 0)\,,\nonumber\\
U(1)_2 & = & \alpha_2(-2,\, 1,\, 1,\, 0,\, \dots,\, 0)\,,\nonumber\\
U(1)_3 & = & \alpha_3(-3,\, 1,\, 1,\, 1,\, \dots,\, 0)\,,\nonumber\\
\vdots &  & \vdots\nonumber\\
\overline{U(1)}_{N-1} & = & \alpha_{N-1}\left(-{N-1 \over N}, \, {1\over N},\, \dots,\, {1 \over N}\right)
\label{groupsUi}
\eea
where $\alpha_i \in [0,2\pi)$ are the angles parametrizing phase rotations.
The next step is to study which subgroups of the maximal torus $T$ can be realizable in the scalar sector of NHDM.

We start from the most general $T$-symmetric potential:
\be
V(T) = - \sum_a m_a^2(\phi_a^\dagger \phi_a) + \sum_{a,b} \lambda_{ab} (\phi_a^\dagger \phi_a)(\phi_b^\dagger \phi_b)
+ \sum_{a \not = b} \lambda'_{ab} (\phi_a^\dagger \phi_b)(\phi_b^\dagger \phi_a)\,,\label{Tsymmetric}
\ee

Each term in this potential transforms trivially under the entire $T$. A sufficiently general potential of this form has no other unitary symmetry. 
It is proven mathematically \cite{Ivanov-Keus-Vdovin} that when we start from the $T$-symmetric potential (\ref{Tsymmetric}) and add more terms, we will never generate any new unitary symmetry that was not already present in $T$. This is the crucial step in proving that the groups described below are realizable.

The Higgs potential is a sum of monomial terms which are linear or quadratic in $\phi_a^\dagger \phi_b$.
Every bilinear $\phi_a^\dagger \phi_b$, $a \neq b$ gains a phase change under $T$
which linearly depends on the angles $\alpha_i$: $\phi_a^\dagger \phi_b \to \exp[i(p_{ab}\alpha_1 + q_{ab}\alpha_2 + \dots + t_{ab}\alpha_{N-1})]\cdot \phi_a^\dagger \phi_b$ with integer coefficients $p_{ab},\, q_{ab},\, \dots,\, t_{ab}$.

Consider a Higgs potential $V$ which, in addition to the $T$-symmetric part (\ref{Tsymmetric}) contains $k\ge 1$ additional terms, with coefficients $p_i,\,q_i,\,\dots t_i$.
This potential defines the following $(N-1)\times k$ matrix of coefficients:
\be
\label{XV}
X(V) = \left(\begin{array}{cccc}
p_1 & q_1 & \cdots & t_1\\
p_2 & q_2 & \cdots & t_2\\
\vdots & \vdots && \vdots \\
p_k & q_k & \cdots & t_k
\end{array}
\right)\,.
\ee
The symmetry group of this potential can be derived from the set of non-trivial solutions for $\alpha_i$ of the following equations:
\be
X(V) \left(\begin{array}{c} \alpha_1 \\ \vdots \\ \alpha_{N-1} \end{array}\right)
= \left(\begin{array}{c} 2\pi n_1 \\ \vdots \\2 \pi n_{k} \end{array}\right)\,,\label{XVeq}
\ee

To find which symmetry groups can be obtained, we need to diagonalize the $X(V)$ matrix with integer entries. After diagonalization, the $X$ matrix becomes $diag(d_1,\dots, d_{N-1})$, where $d_i$ are non-negative integers. The symmetry group of this matrix is $Z_{d_1} \times \cdots \times Z_{d_{N-1}}$
(where $Z_1$ means no non-trivial symmetry, and $Z_0$ means a continuous group $U(1)$).

\section{Abelian symmetries of the 3HDM and 4HDM}

In the 3HDM the representative maximal torus $T \subset PSU(3)$ is parametrized as
\be
T = U(1)_1\times U(1)_2\,,\quad U(1)_1 = \alpha(-1,\,1,\,0)\,,\quad U(1)_2 = \beta\left(-{2 \over 3},\, {1 \over 3},\, {1 \over 3}\right)\,,
\label{3HDM-maximaltorus}
\ee
where $\alpha,\beta \in [0,2\pi)$.
The most general Higgs potential symmetric under this $T$, can be constructed from (\ref{Tsymmetric}).

The full list of subgroups of the maximal torus realizable as the symmetry groups of the Higgs potential in this case:
\be
Z_2,\quad Z_3,\quad  Z_4,\quad Z_2\times Z_2,\quad U(1),\quad U(1)\times Z_2,\quad U(1)\times U(1)\,. \label{list3HDM}
\ee
Writing the explicit potentials that are symmetric under each group in \ref{list3HDM} can be found in \cite{Ivanov-Keus-Vdovin}.

The only Abelian group that is not contained in any maximal torus in $PSU(3)$ is $Z_3 \times Z_3$.
Although there are many such groups inside $PSU(3)$, all of them are conjugate to each other. Thus, only one
representative case can be considered, and we describe it with the following two generators
\be
a = \left(\begin{array}{ccc} 1 & 0 & 0 \\ 0 & \omega & 0 \\ 0 & 0 & \omega^2 \end{array}\right),\quad
b = \left(\begin{array}{ccc} 0 & 1 & 0 \\ 0 & 0 & 1 \\ 1 & 0 & 0 \end{array}\right),\quad \omega = \exp\left({2\pi i \over 3}\right)\,.
\ee
In turns out that the $Z_3 \times Z_3$-symmetric potential is also symmetric under the group $Z_3 \times S_3$, which is non-Abelian. Therefore, we conclude that the symmetry group $Z_3 \times Z_3$ is not realizable for 3HDM.

One could use our strategy to find the full list of realizable Abelian symmetries in 4HDM. However, here we use the results of the next section to find the list of all realizable finite subgroups of the maximal torus in 4HDM:
\be
Z_k\ \mbox{with}\ k = 2,\, \dots ,\, 8; \qquad
Z_2\times Z_k\ \mbox{with}\ k = 2,\,3,\,4; \qquad
Z_2 \times Z_2 \times Z_2\,.\label{list4HDMfinite}
\ee

\section{Abelian symmetries in general NHDM}\label{section-NHDM}

The algorithm described above can be used to find all Abelian groups realizable as the symmetry groups of the Higgs potential for any $N$. We do not yet have the full list of finite Abelian groups for a generic $N$, although we put forth a conjecture concerning this issue. Several strong results can also be proved \cite{Ivanov-Keus-Vdovin} about the order and possible structure of finite realizable subgroups of the maximal torus.

\begin{itemize}
\item{Upper bound on the order of finite Abelian groups}
\\The exact upper bound on the order of the realizable finite subgroup of maximal torus in NHDM is $|G| \le 2^{N-1}$.

\item{Cyclic groups}
\\The cyclic group $Z_p$ is realizable for any positive integer $p \le 2^{N-1}$.

\item{Products of cyclic groups}
\\Let $(N-1) = \sum_{i=1}^k n_i$ be a partitioning of $N-1$ into a sum of non-negative integers $n_i$.
Then, the finite group $G = Z_{p_1}\times Z_{p_2}\times \cdots \times Z_{p_k}\label{manyZn}$ is realizable for any $0 < p_i < 2^{n_i}$.

\item{Conjecture}
\\Any finite Abelian group with order $\le 2^{N-1}$ is realizable in NHDM.

\end{itemize}

\section{Conclusions}\label{section-discussion}

In this work we made a step towards classification of possible symmetries of the scalar sector of the NHDM.
Namely, we studied which Abelian groups can be realized as symmetry groups of the NHDM potential.
We developed an algorithmic strategy that gives full list of possible realizable Abelian symmetries for any given $N$.
We gave the result of our strategy for 3HDM and 4HDM. We also proved that the order of any realizable finite group in NHDM is $\le 2^{N-1}$ and explicitly described which finite groups can appear at given $N$. Finally, we conjectured that {\em any} finite Abelian group with order $\le 2^{N-1}$ is realizable in the NHDM. An obvious direction of future research is to understand phenomenological consequences of the symmetries found. One particular result already obtained \cite{ZpDM} is the possibility to accommodate scalar dark matter stabilized by the $Z_p$ symmetry group, for any $p \leq 2^{N-1}$.


\end{document}